\documentclass{article}

\def\Diracop{\not\negthinspace\partial}
\def\Kostantop{\not\negthinspace\mathcal{K}}
\begin{document}
\title{Kostant's cubic Dirac operator \\ of Lie superalgebras}
\author{Teparksorn Pengpan\footnote{Electronic mail: tpengpan@phys.ufl.edu}}
\date{}
\maketitle

\centerline{\em  Institute for Fundamental Theory,}
\centerline{\em Department of Physics, University of Florida}
\centerline{\em Gainesville FL 32611, USA}

\vskip 2cm

\begin{abstract}
We extend equal rank embedding of reductive Lie algebras to that
of basic Lie superalgebras. The Kac character formulas for equal
rank embedding are derived in terms of subalgebras and Kostant's
cubic Dirac operator for equal rank embedding of Lie superalgebras
is constructed from both even and odd generators and their related
structure constants.
\end{abstract}

\vspace*{\fill}

\noindent PACS number(s): 02.20.Sv, 04.65.+e, 12.60.Jv
\newline

\section{Introduction}

The study of some patterns~\cite{PR}, connected with $N=1$
supergravity theory in eleven dimensions, has led to a recent
understanding in terms of a Weyl character formula by 
GKRS~\cite{GKRS}, based on equal rank embeddings of reductive
Lie algebras. By using the construction of equal rank embedding of
reductive Lie algebras, all possible equal rank
embeddings were cataloged, and nearly all
supersymmetric multiplets of massless and massive particles which
have already known in supersymmetric gauge field theories emerge
as the lowest lines of the infinite tower multiplet spectra, some
of them shown in details in Ref.~\cite{PR}. Immediately after the
appearance of  Weyl character formula for equal rank embedding as
an index formula for Dirac operator, Kostant moved the subject
forward and related the multiplet spectrum to the kernel of a cubic
Dirac operator which he introduced around 30 years
ago~\cite{Kostant}.

In this paper, we extend the Gross-Kostant-Ramond-Sternberg's Weyl
character formula and Kostant's cubic Dirac operator for equal
rank embedding of reductive Lie algebras to those of basic Lie
superalgebras. In section II, we give a brief review in [GKRS]'s
paper. A derivation of the Weyl character formula for an equal rank
embedding is shown. In section III, we extend the equal rank
embedding construction of reductive Lie algebras shown in section
II to that of basic Lie superalgebras. The Kac character formulas
are written in terms of equal rank subalgebras. In section IV, we
give a simple and explicit formulation for a typical
representation of type I Lie superalgebras. In section V, we build
a multiplet of type I Lie superalgebras from that of reductive Lie
algebras. In section VI, Kostant's cubic Dirac operator is
constructed for full Lie superalgebras and then for equal rank
embeddings.

\section{Weyl character formula and equal rank embeddings of
reductive Lie algebras}

Let r be an equal rank subalgebra of reductive Lie algebras, g,
and let \textbf{C} be order of $C$, the ratio of Weyl group of $g$
to that of $r$. The restricted conditions for this kind of equal
rank embedding, $g\supset r$, are that (1) positive roots of $g$
must contain those of $r$, i.e. $\Phi^+(g)\supset \Phi^+(r)$, and
(2) the simple roots of $g$ and $r$ must be chosen consistently so
that the positive Weyl chamber of $r$ contains that of $g$. In
Cartan-Weyl basis, the $c$ elements of Weyl group in $C$ acting on
the sum of highest weight, $\lambda,$ and Weyl vector of $g$,
$\rho_g$, and then after subtracting by Weyl vector of $r$,
$\rho_r$, generate \textbf{C} irreducible representations of $r$,
called \textbf{C}-multiplet, $$c\bullet
\lambda:=c(\lambda+\rho)_g-\rho_r. $$ The Weyl character formula
of the irreducible representation of $g,$ $V_{\lambda},$ can be
rewritten in terms of the irreducible representation of $r,$
$U_{c\bullet \lambda},$ as follows:
\begin{eqnarray}
\hbox{ch}V_{\lambda} & = & \frac {\sum_{w\in W(g)}
\hbox{sgn}(w)e^{w(\lambda+\rho_g)}}{\sum_{w\in W(g)}
\hbox{sgn}(w)e^{w\rho_g}} \nonumber \\ & = & \frac {\sum_{c\in C}
\hbox{sgn}(c)\left(\sum_{w_r\in
W(r)}\hbox{sgn}(w_r)e^{w_r(c\bullet \lambda+\rho_r)}\right)}
{\left(\prod_{\phi\in
\Phi^+(g/r)}(e^{\frac{\phi}{2}}-e^{-\frac{\phi}{2}})\right)
\left(\prod_{\phi\in
\Phi^+(r)}(e^{\frac{\phi}{2}}-e^{-\frac{\phi}{2}})\right)}
\nonumber \\  & = & \frac{1}{\Delta}\sum_{c\in
C}\hbox{sgn}(c)\hbox{ch}U_{c\bullet \lambda}, \label{GKRS1}
\end{eqnarray}
where $\Delta$ is the character difference of two
spinor modules, $S^+$ and $S^-$, of $SO(p=g/r)$, i.e.
$$\Delta:=\prod_{\phi\in
\Phi^+(g/r)}(e^{\frac{\phi}{2}}-e^{-\frac{\phi}{2}})=\hbox{ch
}S^+-\hbox{ch }S^-. $$ The beauty of Eq.(\ref{GKRS1}) is that it
gives us
\begin{equation} V_{\lambda}\otimes
S^+-V_{\lambda}\otimes S^- = \sum_{c\in C}
\hbox{sgn}(c)U_{c\bullet \lambda}.  \label{GKRS2}
\end{equation}
Eq.(\ref{GKRS2}) should be viewed as an equation in the
Grothendieck ring of $r$. The LHS is the character of the
difference between two representations of the same dimension the
representation of $g$ with highest weight $\lambda$, this
representation restricted to $r$ tensored with each of the half
spin representations of $SO(p)$, where $p = \hbox{dim }g -
\hbox{dim }r$, restricted to $r$. In the other words, the LHS is
the algebraic index of the Dirac operator associated to $\lambda$
and the two half spin representations. The alternating sum on the
RHS is just the dimension difference between kernel and cokernel
of the Dirac operator. All the representations on the RHS are
inequivalent and so the RHS  is the end result of a lot of
cancellation in the LHS, in short an index formula for the Dirac
operator. One of the remarkable consequences of Eq.(\ref{GKRS2})
is that, on the LHS, the order of difference of
$V_{\lambda}\otimes S^+$ and $V_{\lambda}\otimes S^-$ is always
equal to \textbf{C}, independent of $\lambda$, and, on the RHS,
the multiplicity of each $U_{c\bullet \lambda}$ representation is
exactly one.

Notice that \textbf{C} is equal to the Euler number which is a
topological invariant of the coset manifold, $G/R,$ corresponding
to exponentiation of $g/r.$ From our catalog of equal rank
embedding of reductive Lie algebras, we would like to give some
examples of the coset manifolds where supersymmetric multiplets
appear to be in their lowest lines of the infinite tower multiplet
spectra. $N=2$ hypermultiplet, $N=4$ vector multiplet, and $N=8$
supergravity multiplet which undoubtedly emerge in the lowest
lines of $SU(N+1)\supset SU(N)\times U(1)$ and $SO(N+2)\supset
SO(N)\times SO(2)$ series live on the N-dimensional complex
projective space and $(N+2)$-dimensional complex Grassmannian
manifold, respectively. If SO(2) or U(1) is viewed as the
light-cone little group, these lowest line spectra are massless
supermultiplets in 4-dimensional space-time. Whereas $N=1,$ $N=2,$
$N=3,$ and $N=4$ massive (massless) multiplets in 4-dimensional
(6-dimensional) space-time which emerge in the lowest lines of
$Sp(2N+2)\supset Sp(2N)\times Sp(2)$ series live on N-dimensional
quaternionic projective space. The last multiplet that we would
like to mention is the $N=1$ massless (massive) supergravity
triplet in 11-dimensional (10-dimensional) space-time. The triplet
emerges from $F_4\supset SO(9)$ and lives on (16-dimensional)
Cayley plane. All infinite tower multiplet spectra are kernel of
Kostant'cubic Dirac operator, $\Kostantop$, $$
\hbox{Ker}(\Kostantop_{\lambda}^2)=\hbox{Ker}(\Kostantop_{\lambda})=\sum_{c\in
C} \hbox{sgn}(c) U_{c\bullet \lambda}. $$ For more analytical
details on Kostant's operator, see Ref.~\cite{BR}.

\section{Kac character formulas and equal rank embedding of basic
Lie superalgebras}

Now, we extend the results of reductive Lie algebras to Lie
superalgebras with non-degenerate Killing form. According to Kac's
classification~\cite{Kac}, there are two types of basic Lie
superalgebras,  type I which is $su(m|n)$ and $osp(2|2n)$ and type
II which is $osp(2m+1|2n),$  $osp(1|2n),$ $osp(2m|2n),$
$osp(4|2;\alpha),$ $F(4),$ and $G(3).$

Let $g=g_{even}\oplus g_{odd}$ be the Lie superalgebras with the
root system $\Phi=\Phi_{even}\cup \Phi_{odd}$. For type I,
$g_{even}$ is simple, i.e. $g_{even}=g_0$, and, for type II,
$g_{even}$ can be graded into $g_2\oplus g_0\oplus g_{-2}$. While,
for the odd part of both type I and II, odd generators can be
graded into fermionic creation and annihilation ones, i.e.
$g_{odd}=g_1\oplus g_{-1}.$ The Poincar\'{e}-Birkhoff-Witt theorem
for Lie algebras can be applied to the case of Lie superalgebras
with some extension~\cite{CNS}. This grading gives us a universal
enveloping algebra, $\mathcal {U}(g),$ e.g., for type I,
$$\mathcal{U}(g)=\mathcal {U}(g_1)\otimes \mathcal {U}(g_0)\otimes
\mathcal {U}(g_{-1}).$$ Define root subsystems, $\overline
\Phi_{even}$ and $\overline \Phi_{odd},$ such that $\overline
\Phi_{even} = \{\alpha\mid \alpha/2 \notin \Phi_{odd}\}$ and
$\overline \Phi_{odd} = \{\beta\mid 2\beta \notin \Phi_{even}\}.$
Since $\Phi_{even}$, $\Phi_{odd}$, $\overline \Phi_{even}$ and
$\overline \Phi_{odd}$ are  invariant under the action of Weyl
group of $g_{even}.$ Hence, Weyl group of $g$ is equal to that of
$g_{even}$, i.e. $W(g) = W(g_{even})$. Define Weyl vector of g to
be one-half the sum of positive even roots minus one-half the sum
of positive odd roots, i.e. $\rho=\rho_{even}-\rho_{odd}.$ Let
$V(\Lambda)$ be a representation of g with $\Lambda$ as a highest
weight in dual Cartan subalgebra. The highest weight
representations of g are classified into typical and atypical. The
representation is typical if, for $\forall \beta \in \overline
\Phi_{odd}^+,$ $(\Lambda + \rho, \beta) \neq 0$; otherwise, it is
atypical. The typical Kac character and supercharacter formulas of
$V(\Lambda)$ are defined, respectively, as
\begin{equation}
\hbox{ch}V(\Lambda)=\frac{N_1}{N_0}\sum_{w\in
W(g)}\hbox{sgn}(w)e^{w(\Lambda+\rho)},   \label{char-Kac1}
\end{equation}
\begin{equation} \hbox{sch}V(\Lambda)=\frac{N_1^{\prime}}{N_0}\sum_{w\in
W(g)}\hbox {sgn}(\overline w)e^{w(\Lambda+\rho)},
\label{schar-Kac1}
\end{equation}
where $$N_0=\prod_{\alpha\in\Phi_{
even}^+}\left(e^{\frac{\alpha}{2}}-e^{-\frac{\alpha}{2}}\right),$$
$$ N_1=\prod_{\beta_{+}\in\Phi_{odd}^+}
\left(e^{\frac{\beta}{2}}+e^{-\frac{\beta}{2}}\right), $$ and $$
N_1^{\prime}=\prod_{\beta_{+}\in\Phi_{odd}^+}
\left(e^{\frac{\beta}{2}}-e^{-\frac{\beta}{2}}\right).$$ The
$\hbox {sgn}(w)$ and $\hbox {sgn}(\overline w)$ are sign change
due to number of reflections with respect to $\Phi_{even}^+$ and
$\overline \Phi_{even}^+.$

In general, in an equal rank embedding of Lie superalgebras $r$ in
$g$ with $\Phi^+(r) \subset \Phi^+(g)$ and \textbf{C} as an index
of the Weyl group of $r$ in the Weyl group of $g$, a
\textbf{C}-multiplet of r is obtained by $$ c\bullet \Lambda :=
c(\Lambda + \rho)_g - \rho_r, $$ where $c \in C$. Under the
condition that  $\Phi_{even,odd}^+(r)$ is invariant under the
action of $c$, i.e. $c\cdot \Phi_{even,odd}^+(r) =
\Phi_{even,odd}^+(r)$, the typical Kac character formula of
$g$-module $V(\Lambda)$ can be written in terms of $r$-module
$U(c\bullet \Lambda)$ as follows:
\begin{eqnarray}
\hbox{ch}V(\Lambda) & = & \frac{\prod_{\beta \in
\Phi_{odd}^+(g)}(e^{\frac {\beta}{2}}+e^{-\frac{\beta}{2}
})}{\prod_{\alpha \in \Phi_{even}^+(g)}(e^{\frac
{\alpha}{2}}-e^{-\frac {\alpha}{2}})} \sum_{w\in W(g)}
\hbox{sgn}(w)e^{w(\Lambda + \rho_g)} \nonumber \\ & = &
\left(\frac{\prod_{\beta \in \Phi_{odd}^+(g/r)}(e^{\frac
{\beta}{2}}+e^{-\frac{\beta}{2} })}{\prod_{\alpha \in
\Phi_{even}^+(g/r)}(e^{\frac {\alpha}{2}}-e^{-\frac
{\alpha}{2}})}\right)\sum_{c\in C} \hbox{sgn}(c)
\hbox{ch}U(c\bullet \Lambda). \label{eqr-char1}
\end{eqnarray}
Similarly done, the typical Kac supercharacter becomes
\begin{equation}
\hbox{sch}V(\Lambda) = \left(\frac{\prod_{\beta
\in \Phi_{odd}^+(g/r)}(e^{\frac {\beta}{2}}-e^{-\frac{\beta}{2}
})}{\prod_{\alpha \in \Phi_{even}^+(g/r)}(e^{\frac
{\alpha}{2}}-e^{-\frac {\alpha}{2}})}\right)\sum_{c\in C}
\hbox{sgn}(c) \hbox{sch}U(c\bullet \Lambda). \label{eqr-schar1}
\end{equation}

For type I Lie superalgebras, there are both typical and atypical
representations. For the typical representation, superdimension,
$\hbox{sdim }V(\Lambda) = \hbox{dim }V_{even}(\Lambda) - \hbox{dim
}V_{odd}(\Lambda)$, is equal to zero. Whereas, $\hbox{sdim
}V(\Lambda)$ of the atypical representation is not. Every type I
odd root is zero-length and $\Phi_{even}^+=\overline
\Phi_{even}^+.$ Since $\rho_{odd}$ is invariant under the action
of Weyl group, i.e. $w\rho_{odd}=\rho_{odd}$. The type I typical
Kac character formula (\ref{char-Kac1}) can be written as

\begin{eqnarray}
\hbox{ch}V(\Lambda) & = &
\prod_{\beta_{+}\in\Phi_{1}^+}(1+e^{-\beta})\hbox{ch}V_0(\Lambda)
\nonumber \\ & = &
\prod_{\beta_{-}\in\Phi_{1}^-}(1+e^{\beta_{-}})\hbox{ch}V_0(\Lambda).
\label{char-Kac2}
\end{eqnarray}
Multiplying out the product
factor on the RHS of Eq.(\ref{char-Kac2}), we obtain a Chern
character of an exterior algebra over $g_{-1},$
$$\prod_{\beta_{-}\in\Phi_{1}^-}(1+e^{\beta_{-}})=
\sum_{n=0}^{n=dim(\Phi_{-1}^{-})}\hbox{ch}(\wedge ^n
g_{-1})=\hbox{ch}(\wedge g_{-1}). $$ So, Eq.(\ref{char-Kac2})
simply becomes
\begin{equation} \hbox {ch}V(\Lambda)=\hbox
{ch}(\wedge g_{-1})\hbox {ch}V_0(\Lambda). \label{char-Kac2.1}
\end{equation}
Similarly, the type I typical Kac supercharacter can be shown
to be
\begin{eqnarray} \hbox{sch}V(\Lambda) & = &
\prod_{\beta_{-}\in\Phi_{1}^-}(1-e^{\beta_{-}})\hbox{ch}V_0(\Lambda)
\nonumber \\ & = & \hbox {sch}(\wedge g_{-1})\hbox
{ch}V_0(\Lambda). \label{schar-Kac2.1}
\end{eqnarray}
On the other
hand, since $\{g_{-1},g_{-1}\}=0,$ the universal enveloping
algebra over $g_{-1},$  $\mathcal {U}(g_{-1}),$ is isomorphic to
the exterior algebra over $g_{-1},$ $\wedge(g_{-1}).$ The g-module
$V(\Lambda)$ can be induced by applying the antisymmetric
combinations of the $g_{-1}$ generators on $V_0(\Lambda),$ i.e.
\begin{equation}
V(\Lambda)= \wedge(g_{-1})\otimes V_0(\Lambda)
\simeq \mathcal {U}(g_{-1})\otimes V_0(\Lambda). \label{V1}
\end{equation}
The character and supercharacter of Eq.(\ref{V1})
are exactly Eq.(\ref{char-Kac2.1}) and Eq.(\ref{schar-Kac2.1}).

In an equal rank embedding, $g\supset r$, of type I which has
\textbf{C} as an index of $W(r)$ in $W(g)$ and is restricted to
the condition that $\Phi^+(g) \supset \Phi^+(r),$
Eq.(\ref{char-Kac2}) becomes $$ \hbox{ch}V(\Lambda) = \frac
1{\Delta}\prod_{\beta_{-}\in\Phi_{1}^{-}(g/r)}
(1+e^{\beta_{-}})\sum_{c\in C}\hbox{sgn}(c)
\left(\prod_{\beta_{-}\in\Phi_{1}^{-}(r)}(1+e^{\beta_{-}})\hbox{ch}U_0(c\bullet
\Lambda)\right), $$ i.e.
\begin{equation}
\hbox{ch}V(\Lambda)\left(\hbox{ch }S^+-\hbox{ch
}S^-\right)=\hbox{ch}\left(\wedge (g_{-1}/r_{-1})\right)
\sum_{c\in C}\hbox{sgn}(c)\hbox{ch}U(c\bullet \Lambda).
\label{char-Kac3}
\end{equation}
Similarly done for the
supercharacter, we obtain
\begin{equation}
\hbox{sch}V(\Lambda)\left(\hbox{ch }S^+-\hbox{ch
}S^-\right)=\hbox{sch}\left(\wedge (g_{-1}/r_{-1})\right)
\sum_{c\in C}\hbox{sgn}(c)\hbox{sch}U(c\bullet \Lambda).
\label{schar-Kac3}
\end{equation}
Eq.(\ref{char-Kac3}) and
Eq.(\ref{schar-Kac3}) correspond to
\begin{equation}
V(\Lambda)\otimes S^+-V(\Lambda)\otimes S^- = \wedge
(g_{-1}/r_{-1})\otimes \sum_{c\in C}  \hbox{sgn}(c)U(c\bullet
\Lambda). \label{V2}
\end{equation}
Now, Eq.(\ref{V2}) can also be
derived explicitly from Eq.(\ref{V1}). By decomposing the $g_{-1}$
basis such that $g_{-1}=(g_{-1}/r_{-1})\oplus r_{-1},$ there
exists a map $(g_{-1}/r_{-1})\oplus r_{-1} \mapsto
(g_{-1}/r_{-1})\otimes 1 + 1\otimes r_{-1}$ from
$(g_{-1}/r_{-1})\oplus r_{-1}$ to $(g_{-1}/r_{-1})\otimes r_{-1}$
which extends uniquely to an isomorphism of exterior algebra, $$
\wedge \left(g_{-1}/r_{-1}\oplus r_{-1}\right) \simeq
\wedge(g_{-1}/r_{-1})\otimes \wedge(r_{-1}). $$ By substituting
the above equation into Eq.(\ref{V1}) and tensoring on both side
by $(S^+-S^-)$, we obtain
\begin{eqnarray*} V(\Lambda)\otimes
S^+-V(\Lambda)\otimes S^- & = & \wedge (g_{-1}/r_{-1})\otimes
\sum_{c\in C}\hbox{sgn}(c)\left( \wedge (r_{-1}) \otimes
U_0(c\bullet \Lambda)\right) \\ & = & \wedge
(g_{-1}/r_{-1})\otimes \sum_{c\in C}\hbox{sgn}(c)U(c\bullet
\Lambda),
\end{eqnarray*}
which is exactly Eq.(\ref{V2}).

For $osp(1|2n)$ of type II Lie superalgebras, every $osp(1|2n)$
odd root has length equal to one-half of the short  positive one
and $$\rho_{osp(1|2n)} = \rho_{even}-\rho_{odd}=
\left(n-\frac12,n-\frac32,\cdots,\frac12\right).$$ Furthermore,
every $osp(1|2n)$ representation is typical, but $\hbox{dim }
V_{even}(\Lambda) \neq \hbox{dim }V_{odd}(\Lambda)$. Nevertheless,
the Kac character and supercharacter formulas of a $osp(1|2n)$
representation can be shown to be similar to that of Lie algebra,
i.e. \begin{equation} \hbox{ch}V(\Lambda)=\frac{\sum_{w\in
W}\hbox{sgn}(w)e^{w(\Lambda+\rho)}} {\sum_{w\in
W}\hbox{sgn}(w)e^{w(\rho)}}, \label{char-Kac4} \end{equation}
\begin{equation}
\hbox{sch}V(\Lambda)=\frac{\sum_{w\in
W}\hbox{sgn}(\overline w)e^{w(\Lambda+\rho)}} {\sum_{w\in
W}\hbox{sgn}(\overline w)e^{w(\rho)}}.
\label{schar-Kac4}
\end{equation}
The equal rank embedding of $osp(1|2m)\times osp(1|2n-2m)$ in
$osp(1|2n)$ is the only possible type with the full
subsuperalgebra. In this case, all odd generators of g are
completely eaten by the r. Whereas, the even generators of g which
are left form the orthogonal complement basis to the even basis of
r under the killing form of g. Under the restriction to r-module,
Eq.(\ref{char-Kac4}) and Eq.(\ref{schar-Kac4}) simply become
\begin{equation}
\hbox{ch}V(\Lambda)=\frac{1}{\Delta}\sum_{c\in
C}\hbox{sgn}(c)\hbox{ch}U(c\bullet \Lambda),
\label{char-Kac5}
\end{equation}
and
\begin{equation}
\hbox{sch}V(\Lambda)=\frac{1}{\Delta}\sum_{c\in
C}\hbox{sgn}(c)\hbox{sch}U(c\bullet \Lambda). \label{schar-Kac5}
\end{equation}
Both Eq.(\ref{char-Kac5}) and Eq.(\ref{schar-Kac5})
correspond to
\begin{equation} V(\Lambda)\otimes
S^+-V(\Lambda)\otimes S^- = \sum_{c\in C} \hbox{sgn}(c)U(c\bullet
\Lambda). \label{V3}
\end{equation}

For the rest of type II, we use Kac character and supercharacter
formulas of equal rank embedding, Eq.(\ref{eqr-char1}) and
Eq.(\ref{eqr-schar1}).  If $g_1$ and $g_{-1}$ are vector spaces of
odd generators, there is a canonical linear map from $\wedge^a g_1
\otimes \wedge^b g_{-1}$ to $\wedge^{a+b} (g_1 \oplus g_{-1})$,
which takes $((g_1)_1 \wedge \cdots \wedge (g_1)_a)\otimes
((g_{-1})_1 \wedge \cdots \wedge (g_{-1})_b)$ to $((g_1)_1 \wedge
\cdots \wedge (g_1)_a\wedge (g_{-1})_1 \wedge \cdots \wedge
(g_{-1})_b)$. This determines a linear isomorphism

\begin{eqnarray*}
\wedge (g_1 \oplus g_{-1}) & \simeq & \bigoplus_{a=0}^N
\left(\wedge^{N-a} g_1\otimes \wedge^{a} g_{-1}\right) \\ & \simeq
& \wedge g_1 \otimes \wedge g_{-1} .
\end{eqnarray*} The prefactor
$N_1$ of Kac character formula is generally the character of the
exterior algebra over the direct sum of $g_1$ and$g_{-1}$ vector
spaces
\begin{eqnarray*} \prod_{\beta\in \Phi_1^+}\left(e^{\frac \beta 2} +
e^{-\frac \beta 2}\right) & = & \prod_{\beta_{\pm}\in
\Phi_1^{\pm}}\left(e^{\frac {\beta_+} 2} + e^{\frac {\beta_-}
2}\right)  \\ & = & \hbox{ch}\wedge( g_1 \oplus g_{-1}).
\end{eqnarray*}
Finally, the type II typical Kac character and
supercharacter can be written as
\begin{equation}
\hbox{ch}V(\Lambda)\left(\hbox{ch }S^+ - \hbox{ch }S^-\right) =
\hbox{ch}\wedge( g_1/r_1 \oplus g_{-1}/r_{-1})\sum_{c\in
C}\hbox{sgn}(c)\hbox{ch}U(c\bullet \Lambda), \label{char-Kac6}
\end{equation}
and
\begin{equation}
\hbox{sch}V(\Lambda)\left(\hbox{ch }S^+ - \hbox{ch }S^-\right) =
\hbox{sch}\wedge (g_1/r_1 \oplus g_{-1}/r_{-1}) \sum_{c\in
C}\hbox{sgn}(c)\hbox{sch}U(c\bullet \Lambda), \label{schar-Kac6}
\end{equation}
which correspond to
\begin{equation} V(\Lambda)\otimes
S^+-V(\Lambda)\otimes S^- = \wedge (g_1/r_1 \oplus
g_{-1}/r_{-1})\otimes \sum_{c\in C} \hbox{sgn}(c)U(c\bullet
\Lambda). \label{V4}
\end{equation}

\section{Representations of $\wedge g_{-1}$ of type I Lie
superalgebras}

For type I Lie superalgebras with non-degenerate Killing form, the
typical representation is induced by applying $g_{-1}$
generators on $g_0$-module. Therefore, we need to know an explicit
representation of $\wedge g_{-1}$ in terms of $g_0$-module. Let N be
dimension of $\Phi_{1}^{-}$. The exterior algebra over $g_{-1}$ is
$$ \wedge g_{-1} =  \bigoplus_{k=0}^{N} \wedge^{k}g_{-1}, $$
with dimension,
$$
 \hbox{dim} (\wedge g_{-1}) =  \bigoplus_{k=0}^{N} \hbox{dim}
(\wedge^{k} g_{-1}) = \sum_{k=0}^{N} \left( \begin{array}{c} N \\ k
\end{array} \right) = 2^N,$$
and superdimension,
$$
\hbox{sdim} (\wedge g_{-1}) =  \bigoplus_{k=0}^{N} \hbox{sdim}
(\wedge^{k} g_{-1}) = \sum_{k=0}^{N}(-1)^k \left( \begin{array}{c} N \\ k
\end{array} \right) = 0.
$$ Let $Q_{i=1,2,\dots,N}^{\dagger}$ be N completely antisymmetric
fermionic generators which transform in the fundamental
representation of $g_0$. The fermionic generators generate even
and odd modules which are isomorphic to a direct sum of two spinor
representations of $so(2N)$. One of spinor representation of
$so(2N)$ is the even module of $\wedge g_{-1}$, called bosonic
module, and the other is the odd module of $\wedge g_{-1}$, called
fermionic module. Let $T^+$ be the bosonic module and $T^-$ be the
fermionic module of $so(2N)$ such that \begin{eqnarray*}
 T^+ & = & \wedge^0 g_{-1} \oplus \wedge^2 g_{-1} \oplus \wedge^4 g_{-1}
\oplus \dots \\ & = & \left((Q_i^\dagger)^0\oplus (Q_i^\dagger)^2
\oplus (Q_i^\dagger)^4\ \oplus \dots \space \right)|1>_0 \\ &
\equiv & \hbox{bosonic module}, \end{eqnarray*} and
\begin{eqnarray*}
 T^- & = & \wedge^1 g_{-1} \oplus \wedge^3 g_{-1} \oplus \wedge^5 g_{-1}
\oplus \dots \\ & = & \left((Q_i^\dagger)^1\oplus (Q_i^\dagger)^3
\oplus (Q_i^\dagger)^5\ \oplus \dots \space \right)|1>_0 \\ &
\equiv & \hbox{fermionic module}. \end{eqnarray*} With restriction
to $g_0$-module, the type I typical representation is
\begin{eqnarray*} V(\Lambda) & = & \wedge g_{-1}\otimes
V_0(\Lambda) \\ & = & \left(T^{+} \oplus T^{-}\right)\otimes
V_0(\Lambda). \end{eqnarray*}

For $su(m|n)$, $\wedge g_{-1}$ representation is isomorphic to a direct sum of
two spinor representations of
$so(2mn)$. With restriction to $su(m)\times su(n)\times u(1)$,
$Q_i^\dagger$ transforms as $(m,n)_{-1}$, where -1 is a $u(1)$
charge.
\newline \newline
\textbf{Ex.1} $su(2|1)$
\newline
$$\hbox{dim}(g_{-1}) = 2^2 = 1 + 2 + 1$$
$Q_i^\dagger \sim 2_{-1}$
\begin{eqnarray*}
T^+ & = & 1_0 \oplus 1_{-2} \\
T^{-} & = & 2_{-1}
\end{eqnarray*}
$$V_{su(2|1)}(\Lambda)=(T^+ \oplus T^-)\otimes V_{su(2)\times u(1) }(\Lambda)$$
\newline
\textbf{Ex.2} $su(3|1)$
\newline
$$\hbox{dim} (g_{-1}) = 2^3 = 1 + 3 + 3 + 1$$
$Q_i^\dagger \sim 3_{-1}$
\begin{eqnarray*}
T^+ & = & 1_0 \oplus \overline{3}_{-2} \\
T^{-} & = & 3_{-1} \oplus 1_{-3}
\end{eqnarray*}
$$V_{su(3|1)}(\Lambda)=(T^+ \oplus T^-)\otimes V_{su(3)\times u(1)}(\Lambda)$$
\newline
\textbf{Ex.3} $su(3|2)$
\newline
$$\hbox{dim} (g_{-1}) = 2^6 = 1 + 6 + 15 + 20 + 15 + 6 + 1$$
$Q_i^\dagger \sim (3,2)_{-1}$
\begin{eqnarray*}
T^+ & = & (1,1)_0 \oplus (\overline{3},3)_{-2} \oplus (6,1)_{-2} \oplus (6,1)_{-4}
\oplus (3,3)_{-4} \oplus (1,1)_{-6}\\
T^{-} & = & (3,2)_{-1} \oplus (8,2)_{-3} \oplus (4,1)_{-3} \oplus (\overline{3},2)_{-5}
\end{eqnarray*}
$$V_{su(3|2)}(\Lambda)=(T^+ \oplus T^-)\otimes V_{su(3)\times su(2)\times u(1)}(\Lambda)$$
\newline
\textbf{Ex.4} $su(4|2)$
\newline
$$\hbox{dim} (g_{-1}) = 2^8 = 1 + 8 + 28 + 56 +70 + 56 + 28 + 8 +
1$$ $Q_i^\dagger \sim (4,2)_{-1}$ \begin{eqnarray*} T^+ & = &
(1,1)_0 \oplus (10,1)_{-2} \oplus (6,3)_{-2} \oplus (20^{\prime},1)_{-4}
\oplus (15,3)_{-4} \\  & & \oplus (5,1)_{-4} \oplus
(\overline{10},1)_{-6} \oplus (6,3)_{-6} \oplus (1,1)_{-8} \\
T^{-} & = & (4,2)_{-1} \oplus (20,1)_{-3} \oplus
(\overline{4},4)_{-3} \oplus (\overline{20},1)_{-5} \oplus
(4,4)_{-5} \oplus (\overline{4},2)_{-7} \end{eqnarray*}
$$V_{su(4|2)}(\Lambda)=(T^+ \oplus T^-)\otimes V_{su(4)\times
su(2)\times u(1)}(\Lambda)$$

For $osp(2|2n)$, $\wedge g_{-1}$ representation is a direct sum of
two spinor representations of $so(4n)$. With restriction to
$sp(2n)\times u(1)$, $Q_i^\dagger$ transforms as $(2n)_{-1}$ with
-1 as a $u(1)$ charge.  \newline \textbf{Ex.5} $osp(2|4)$
\newline
$$\hbox{dim} (g_{-1}) = 2^4 = 1 + 4 + 6 + 4 + 1$$
$Q_i^\dagger   \sim 4_{-1}$
\begin{eqnarray*}
T^+ & = & 1_0 \oplus 5_{-2} \oplus 1_{-2} \oplus 1_{-4} \\
T^{-} & = &  4_{-1} \oplus 4_{-3}
\end{eqnarray*}
$$V_{osp(2|4)}(\Lambda)=(T^+ \oplus T^-)\otimes V_{sp(4)\times u(1)}(\Lambda)$$
\newline
\textbf{Ex.6} $osp(2|6)$
\newline
$$\hbox{dim} (g_{-1}) = 2^6 = 1 + 6 + 15 + 20 + 15 + 6 + 1$$
$Q_i^\dagger  \sim 6_{-1}$ \begin{eqnarray*} T^+ & = &  1_0 \oplus
14_{-2} \oplus 1_{-2} \oplus 14_{-4} \oplus 1_{-4} \oplus 1_{-6}\\
T^{-} & = &  6_{-1} \oplus 14_{-3}^{\prime} \oplus 6_{-3} \oplus
6_{-5} \end{eqnarray*} $$V_{osp(2|6)}(\Lambda)=(T^+ \oplus
T^-)\otimes V_{sp(6)\times u(1)}(\Lambda)$$

\section{Building type I C-multiplets and Kostant's cubic Dirac
operator}

For type I Lie superalgebras, there is a remarkable point we would
like to mention. From the multiplet spectrum of equal rank
embedding of reductive Lie algebras, $r_0 \subset g_0$, we can
build the spectrum of the type I Lie superalgebras on top of them
by simply tensoring them with a module generated by $r_{-1}$
generators. The $r_{-1}$ generators transform in the fundamental
representation of $r_0$.

Define $\wedge (r_{-1}) = R^+ \oplus R^-$ such that $$ U(\Lambda)
= \wedge (r_{-1})\otimes U_0(\Lambda) = (R^+ \oplus R^-)\otimes
U_0(\Lambda). $$ Tensoring on both side of Eq.(\ref{GKRS2}) with
$\wedge (r_{-1})$, we obtain
\begin{eqnarray}
\wedge (r_{-1})\otimes \left(V_0(\Lambda)\otimes
S^+-V_0(\Lambda)\otimes S^-\right) & = & \sum_{c\in C}
\hbox{sgn}(c)\left(\wedge (r_{-1})\otimes U_0(c\bullet
\Lambda)\right) \nonumber \\ & = & \sum_{c\in C}
\hbox{sgn}(c)U(c\bullet \Lambda). \label{C1}
\end{eqnarray}
Under restriction to $r_0$-module, whenever $V_0(\Lambda)$ on the
LHS of Eq.(\ref{C1}) is $su(n)\times u(1)$, or $sp(2n)\times
u(1)$, there is an emergence of type I typical
\textbf{C-}multiplet  on the RHS. Notice in the case that
$U(c\bullet \Lambda)$ is $su(m|n),$ the representations of
$R^{\pm}$ are similar to those of $S^{\pm}$ except $u(1)$ values.

Let $g = r \oplus p$ be Lie superalgebras where $p$ is the
orthogonal complement to $r$ under the non-degenerate Killing form
of $g$. Let $p = p_{even} \oplus p_{odd} = p_0 \oplus p_1 \oplus
p_{-1}$. Tensoring on both sides of Eq.(\ref{C1}) by $\wedge
p_{-1}$, we get
\begin{equation}
V(\Lambda)\otimes S^+-V(\Lambda)\otimes S^- =  \wedge
p_{-1}\otimes \sum_{c\in C}  \hbox{sgn}(c)U(c\bullet \Lambda),
\label{C2}
\end{equation}
which is exactly Eq.(\ref{V2}). Now, we need to get rid of $\wedge
p_{-1}$ on the RHS of  Eq.(\ref{C2}) by mapping it into identity.
To do so, we do a tensor product on both sides of Eq.(\ref{C2}) by
the following contraction , sometimes called an internal product,
on exterior powers between vector space and its dual: $$
\wedge^N(p_{odd}) = \wedge^N p_1 \otimes \wedge^N p_{-1} =
\mathbf{1}.$$
Eq.(\ref{C2}) becomes
\begin{equation}
(\wedge^N p_{odd})\otimes V(\Lambda)\otimes (S^+ - S^-) =
\mathbf{1}\otimes \sum_{c\in C} \hbox{sgn}(c)U(c\bullet \Lambda).
\label{C3}
\end{equation}

As already known, the character form of a Dirac operator,
$\Diracop \in C(p)$, with a map $\Diracop:S^\pm \rightarrow S^\mp$
is $$ \hbox {ch}(\Diracop) = \hbox{ch}(S^+) - \hbox{ch}(S^-). $$
The character of Eq.(\ref{GKRS1}) implies that there exists a
Kostant's Dirac operator, $\Kostantop\in
\mathcal{U}(g_{even})\otimes C(p_{even})$, with a map
\begin{equation}
\Kostantop_{\lambda}:V_{\lambda} \otimes S^\pm \rightarrow
V_{\lambda} \otimes S^\mp.
\end{equation}
Similarly, the character of Eq.(\ref{C3}) suggests that there
exists the operator for Lie superalgebras,
$\Kostantop_{\Lambda}\in \mathcal{U}(p_{odd})\otimes
\mathcal{U}(g_{even}\oplus g_{odd})\otimes C(p_{even})$, with a
map
\begin{equation}
\Kostantop_{\Lambda}:V(\Lambda) \otimes S^\pm \rightarrow
V(\Lambda) \otimes S^\mp.
\end{equation}

\section{Kostant's cubic Dirac operator for an equal rank
embedding of Lie superalgebras}

Let $L_i$ and $F_a$ be even and odd generators, respectively, for
$Z_2$-graded Lie superalgebras such that $$[L_i,L_j] =
f_{[ijk]}L_k,$$ $$\{F_a,F_b\} = f_{i(ab)}L_i,$$ and $$[L_i,F_a] =
f_{[ia]b}F_b,$$ where i,j, k = 1, 2, \dots, dim($g_{0}$) and a, b
= 1, 2, \dots, dim($g_{1}$). The Kostant's cubic Dirac operator of
Lie superalgebras is extended from that of Lie algebras by adding
just two terms, a linear term in odd operators and a structure
constant term, i.e. \begin{equation} \Kostantop_{\Lambda} =
\Kostantop_{\Lambda}^{0} + \Kostantop_{\Lambda}^{1},
\label{Kostant01} \end{equation} where \begin{equation}
\Kostantop_{\Lambda}^{0} = \gamma_iL_i -
\frac{1}{2}\gamma_{[ijk]}f_{[ijk]} \label{Kostant02}
\end{equation} and
\begin{equation} \Kostantop_{\Lambda}^{1} = \alpha_aF_a -
\frac{1}{2}\gamma_i\alpha_{(ab)}f_{i(ab)} \label{Kostant03}
\end{equation} Sum over all indices is assumed in the above
equations, where [\dots] in the subscript is for antisymmetric sum
and (\dots) for symmetric sum. Eq.(\ref{Kostant02}) is the
Kostant'cubic Dirac operator for reductive Lie algebras. The
$\gamma$-matrices associated to even generators are normalized so
that $$\{\gamma_i,\gamma_j\} = \delta_{ij},$$ which gives
$$\{\gamma_{i^{\prime}},\gamma_{[ijk]}\} =
\delta_{i^{\prime}k}\gamma_{[ij]}.$$ The $\alpha$-matrices
 associated to odd generators are subjected to the
following conditions.
 \newline
(I) $\{\gamma_i,\alpha_a\} = 0$. This relation is consistent with
the antisymmetric property of product of even and odd generators.
\newline
(II) $\{\alpha_a,[\alpha_b,\alpha_c]\} +
\{[\alpha_a,\alpha_c],\alpha_b\} =
 [\alpha_{(ab)},\alpha_c]= 0.$ This property is due to invariance
 of odd gennerators under Killing form.

Squaring Eq.(\ref{Kostant02}), we get \begin{eqnarray}
(\Kostantop_{\Lambda}^{0})^2 & = &
\frac{1}{2}\gamma_{(ij)}\{L_i,L_j\} +
\frac{1}{2}\gamma_{[ij]}[L_i,L_j] \nonumber \\ & & -
\frac{1}{2}\{\gamma_{i^{\prime}},\gamma_{[ijk]}\}f_{[ijk]}L_{i^{\prime}}
+ \left( \frac{1}{2}\gamma_{[ijk]}f_{[ijk]} \right)^2  \nonumber
\\ & = & \frac{1}{2}\gamma_{(ij)}\{L_i,L_j\} + \left(
\frac{1}{2}\gamma_{[ijk]}f_{[ijk]} \right)^2 \label{Kostant2e}
\end{eqnarray} Notice that the linear terms in even generators cancel each
other. Thus, Eq.(\ref{Kostant2e}) is explicity invaraint under the
action of even and odd generators. The first term of
Eq.(\ref{Kostant2e}) is the quadratic Casimir operator of
reductive Lie algebras, \begin{equation} C_2^{0}(\Lambda) =
\frac{1}{2}\gamma_{(ij)}\{L_i,L_j\}. \end{equation} Since
$\rho_{0}$, one-half the sum of positive even roots, can be
identified as
\begin{equation} \rho_{0} = \frac 12\gamma_{[ijk]}f_{[ijk]}. \end{equation}
The second term of Eq.(\ref{Kostant2e}) is the Freudenthal-de
Vries' strange formula, \begin{eqnarray} (\rho,\rho)_{0} & = &
\left(\frac{1}{2}\gamma_{[ijk]}f_{[ijk]} \right)^2 =
\frac{1}{24}dim(g_{0})h_{0}^{\vee}(\theta,\theta)_{0} \nonumber
\\ & = & \frac{1}{24}dim(g_{0})C_2^{0,ad}.
\end{eqnarray} Where, in the above equation, $h^{\vee}$ is dual Coxeter
number, $\theta$ is the highest root, and $C_2^{ad}$ is the
quadratic Casimir value in the adjoint representation.

Squaring Eq.(\ref{Kostant03}), we get \begin{eqnarray}
(\Kostantop_{\Lambda}^{1})^2 & = & \frac 12\alpha_{[ab]}[F_a,F_b]
+ \frac 12 \alpha_{(ab)}\{F_a,F_b\} \nonumber \\ & & - \frac 12
\gamma_i[\alpha_{(ab)},\alpha_{a^{\prime}}]f_{i[ab]}F_{a^{\prime}}
+ \left(\frac{1}{2}\gamma_i\alpha_{(ab)}f_{i(ab)} \right)^2
\nonumber \\ & = & \frac{1}{2}\alpha_{[ab]}[F_a,F_b] +
\frac{1}{2}\alpha_{(ab)}f_{i(ab)}L_i
 + \left( \frac{1}{2}\gamma_i\alpha_{(ab)}f_{i(ab)} \right)^2.
\label{Kostant2o} \end{eqnarray} In contrary to the even part,
Eq.(\ref{Kostant2o}) by itself is not invariant due to the
presence of a linear term in even generators. The linear term in
$L_i$ will be cancelled out by one of the the cross terms of the
square of the combined even and odd Kostant's cubic Dirac
operator, \begin{eqnarray} (\Kostantop_{\Lambda})^2 & = &
(\Kostantop_{\Lambda}^{0})^2 + (\Kostantop_{\Lambda}^{1})^2 +
\{\gamma_i,\alpha_a\}L_iF_a -
\frac{1}{2}\{\gamma_{[ijk]},\alpha_a\}f_{[ijk]}F_a \nonumber \\ &
& - \frac 12
\{\gamma_i,\gamma_{i^\prime}\}\alpha_{(ab)}f_{i^{\prime}(ab)}L_i
 - \frac 12 \gamma_i[\alpha_{(ab)},\alpha_{a^\prime}]f_{i(ab)}F_{a^\prime} \nonumber \\
& & + \frac 12
\{\gamma_{[ijk]},\gamma_{i^\prime}\}\alpha_{(ab)}f_{[ijk]}f_{i^\prime(ab)}
\nonumber \\ & = & \gamma_{(ij)}L_{(i}L_{j)} +
\alpha_{[ab]}F_{[a}F_{b]} +
\left(\frac{1}{2}\gamma_{[ijk]}f_{[ijk]} \right)^2 \nonumber \\ &
& + \left( \frac{1}{2}\gamma_i\alpha_{(ab)}f_{i(ab)} \right)^2
 + \frac 12 \gamma_{[ij]}\alpha_{(ab)}f_{[ijk]}f_{k(ab)}
\label{Kostant3} \end{eqnarray} Since $\rho_{1}$, one-half the sum
of positive odd roots, can be identified as \begin{equation}
\rho_{1} = - \frac 12 \gamma_i\alpha_{(ab)}f_{i(ab)}.
\end{equation} Recall that, for Lie superalgebras, $\rho =
\rho_{0} - \rho_{1}$. Eq.(\ref{Kostant3}) can be simply written as
\begin{equation} (\Kostantop_{\Lambda})^2 = C_2(\Lambda) +
(\rho,\rho),
\end{equation} where $$ C_2(\Lambda) = \gamma_{(ij)}L_{(i}L_{j)} +
\alpha_{[ab]}F_{[a}F_{b]}. $$ The Laplacian operator turns out to
be invariant under the action of even and odd generators of g
again. For Lie superalgebras, the generalization of Freudenthal-de
Vries strange formula still holds~\cite{KW}
\begin{equation}
(\rho,\rho) = \frac {h^{\vee}}{24}(\hbox{dim }g_0 - \hbox{dim
}g_1),
\end{equation}
where $h^{\vee}$'s, the dual Coxeter numbers, of $g$ are given in
Table 1.

In an equal rank embedding of Lie superalgebras, $g \supset r$,
with $\Phi_g \supset \Phi_r$, let the even and odd generators,
$L_i$ and $F_a$, span the basis of $g$. In according to the $g = r
\oplus p$ decomposition, $L_{\hat{i}}$ and $F_{\hat{a}}$ span the
basis of $r$ and $L_{I}$ and $F_{A}$ span the basis of $p$, the
orthogonal complement of $r$ under the Killing form of g. The
Kostant's cubic Dirac operator on $p = g/r$ is \begin{eqnarray}
\Kostantop_p & = & \Kostantop_g - \Kostantop_r \nonumber \\ & = &
\gamma_{I}L_{I} + \alpha_{A}F_{A} - \frac 12
\gamma_{[IJK]}f_{[IJK]} - \frac 12 \gamma_I\alpha_{(AB)}f_{I(AB)},
\end{eqnarray}
where $f_{[IJK]}$ and $f_{I(AB)}$ are the structure constants of
$g$ that are not in $r$. Since, under the Killing form of g, $r$
and $p$ basis are orthogonal to each other,
$$\{\gamma_{\hat{i}},\gamma_I\} = \{\gamma_{\hat{i}},\alpha_A\} =
\{\alpha_{\hat{a}},\gamma_I\} = \{\alpha_{\hat{a}},\alpha_A\} =
0.$$ As a result, we have $$\{\Kostantop_r,\Kostantop_p\} = 0.$$
The square of Kostant's coset cubic Dirac operator simply is
\begin{eqnarray} (\Kostantop_p)^2 & = & (\Kostantop_g)^2 -
(\Kostantop_r)^2 \nonumber \\ & = & \left( C_2 +
(\rho,\rho)\right)_g - \left(C_2 + (\rho,\rho)\right)_r.
\end{eqnarray}
Notice that both $(\Kostantop_g)^2$ and $(\Kostantop_r)^2$ commute
with the generators of r. Hence, $(\Kostantop_p)^2$ is also
invariant under the action of r.

In conclusion, we have derived the Kac character formulas and have
constructed Kostant's cubic Dirac operator for equal rank
embeddings of Lie superalgebras. In case of reductive Lie
algebras, the coset space method of equal rank embeddings led to a
Kazama-Suzuki model construction of a new class of unitary $N=2$
superconformal theories~\cite{KS} and a subclass of the
construction could be represented by Landau-Ginzburg
models~\cite{LVW}. In case of Lie superalgebras, it deserves to be
pursued whether there will be any relevance of the coset
superspace method to a construction of any physical model.

\section*{Acknowledgements}

The author would like to thank Prof. P. Ramond for invaluable
supervision and Prof. S. Sternberg for pointing out
Ref.~\cite{CNS} and Ref.~\cite{KW} into attention. This work is
supported by Ministry of University Affairs, Thailand.

\newpage

\centerline{\begin{tabular}{|c|c|}
  \hline
  $g$ & $h^{\vee}$ \\
  \hline
  $su(m|n)$ & $|m-n|$ \\
  $osp(2|2n)$ & $n$ \\
  $osp(2m+1|n)$ &
    $2(m-n)-1$ if $m>n$,
    $n-m+\frac 12$ if $m\leq n$  \\
  $osp(2m|n)$ &
    $2(m-n-1)$ if $m\geq n$,
    $n-m+1$ if $m<n+1$   \\
  $osp(4|2;\alpha)$ & 0 \\
  $F(4)$ & 3 \\
  $G(3)$ & 2 \\ \hline
\end{tabular}}
\vskip 0.5cm \centerline{Table 1: Dual Coxeter Numbers of Basic
Lie Superalgebras}

\end{document}